\documentstyle[mahyd209]{article}

\title{The New Mexico alpha-omega Dynamo Experiment:  Modeling
Astrophysical Dynamos}

\author{S. A. Colgate\inst{1}\inst{,2},  V. I. Pariev\inst{2}\inst{,3},
H. F. Beckley\inst{1},    R. Ferrel\inst{1}, \\  V. D.
Romero\inst{1}, J. C. Weatherall\inst{1}}

\institute{Department of Physics,  New Mexico Institute of Mining and
Technology, NM 87801, USA
\and Theoretical Astrophysics Group, T-6, MS B288, Los Alamos National
Laboratory, Los Alamos, NM 87545, USAconstruct
\and P. N. Lebedev Physical Institute, Leninsky Prospect 53, Moscow
117924, Russia}

\begin{document}
\maketitle


\begin{abstract} A magnetic dynamo experiment is under construction at
the New Mexico Institute of Mining and Technology. The experiment is
designed to demonstrate in the laboratory the $\alpha\omega$ magnetic
dynamo, which is believed to operate in many rotating and conducting
astrophysical objects. The experiment uses the Couette flow of liquid
sodium between two cylinders rotating with different angular
velocities to model the $\omega$--effect. The $\alpha$--effect is
created by the rising and expanding jets of liquid sodium driven
through a pair of orifices in the end plates of the cylindrical
vessel, presumably simulating plumes driven by buoyancy in
astrophysical objects. The water analog of the dynamo device has been
constructed and the flow necessary for the dynamo has been
demonstrated. Results of the numerical simulations of the kinematic
dynamo are presented. The toroidal field produced by the
$\omega$--effect is predicted to be $B_{\phi} \simeq (R_m/2\pi)
B_{poloidal}\simeq 20 \times B_{poloidal}$ for the expected magnetic
Reynolds number of $R_m \sim 120$. The critical rate of jets necessary
for the dynamo self-excitation is predicted from the calculations to
be a pair of plumes every 4 revolutions of the outer cylinder.  For
reasonable technical limitations on the strength of materials and the
power of the drive, the self-excitation of the dynamo appears to be
feasible.
\end{abstract}


\section*{Introduction}\label{sec_intro}

Recent years have been marked by exciting developments in the field of
MHD dynamos, in particular the experimental realization of homogeneous
dynamos was achieved.  After many years of research and preparations,
an exponentially growing dynamo mode was observed in the experiment
conducted in a liquid-sodium facility in Riga, Latvia
\cite{gailitis00a,gailitis01}. This experiment reproduces the simplest
dynamo flow proposed by \cite{parker55} and also subsequently by
\cite{ponomarenko73} in the laboratory.  Another successful dynamo
experiment of a different type was built in Karlsruhe, Germany. This
experiment verifies the ability of a regular spatial arrangement of
vortices to amplify the magnetic field. The growth of the magnetic
field starting from the initial seed value of $\approx
1\,\mbox{Gauss}$ up to $\approx 70\,\mbox{Gauss}$ was observed in
Karlsruhe experiment \cite{stieglitz01}. The magnetic field reached
the back-reaction, saturated limit, and the excitation of the
non-axisymmetric mode predicted by the theory was observed. There are a
number of other dynamo experiments, which are under preparation or
discussion. Each of these experiments is designed to test different
flow patterns capable of dynamo action. All these experiments (except
Karlsruhe experiment) are designed to use axisymmetric rotating flows,
either stationary or non-stationary.  We designed another kind of
dynamo experiment, which will use essentially non-axisymmetric,
non-stationary flows.

Here, because the flow is non-stationary and non-axisymmetric, then in
both cases the field averaged over many plume ejections will approach
a near steady state of axisymmetric symmetry.  The flux from such a
dynamo thus may simulate the astrophysical fluxes observed on large
scales.

Real dynamos operate in Nature on astrophysical scales, in planets,
convective envelopes of stars, galactic discs, accretion discs, and,
possibly, on the largest scale in the clusters of galaxies.  
Parker \cite{parker55}, suggested  the $\alpha\omega$ dynamo as an
explanation of the large scale fields of many astrophysical dynamos.
Here,  cyclonic or anticyclonic plumes are the primary source of the
helicity. In the case of planetary and stellar dynamos such plumes are
believed to be rising and sinking convective cells
\cite{parker79,mestel99}.  Here we invoke for both the experiment and the theory of the
accretion disk dynamo, anticyclonic plumes because of their finite
rotation angle, $\simeq \pi/2$ radians.

   In the case of Galactic dynamos, supernovae explosions can also
produce the necessary anticyclonic motions
\cite{ferriere92,ferriere93a,ferriere93b}. Finally, the most energetic
dynamos in the present day universe should exist in the accretion
discs around black holes in the nuclei of active galaxies
\cite{zeldovich83}. The magnetic dynamo in the black hole accretion
discs in the centers of galaxies could be due the stars passing
through the accretion disc and producing the plumes of heated
gas. These plumes rise above the disc plane, expand and produce
anticyclonic motion necessary for the operation of an $\alpha\omega$
dynamo \cite{colgate01,pariev01}. The New Mexico dynamo experiment is
designed to demonstrate the excitation of the astrophysical-type
dynamo due to non-axisymmetric plumes rising through the differentially
rotating liquid sodium.  The design and initial construction phase of
this experiment is underway at the New Mexico Institute of Mining and
Technology located in Socorro (USA). Those involved in the conceptual
design, engineering, mechanical design, and theoretical considerations
of the experiment are associated with the New Mexico Institute of
Mining and Technology and Los Alamos National Laboratory both located
in the State of New Mexico, and so we refer to the experiment as the
New Mexico Dynamo Experiment.

The experiment was designed after a water visualization experiment
confirmed the existence of the two fundamental fluid motions necessary
for a laboratory experiment to demonstrate the $\alpha\omega$ dynamo.
These are: (1) the maximum differential shear and stability of Couette
flow and (2) the production of axially aligned diverging plumes from
pulsed jets and their subsequent anticyclonic and limited rotation
relative to a rotating frame. Far more accurate measurements will be
possible of both these flows in the completed dynamo apparatus, but we
include here the qualitative confirmation of these fundamental flows
as a basis of the design.  In the Theory section we will show how
these two orthogonal flows make an $\alpha\omega$ dynamo when produced
in a conducting fluid of sufficiently large magnetic Reynolds number.

The plan of this paper is to briefly discuss how such an
$\alpha\omega$ dynamo works in Sec.~2 and then in Sec.~3 to briefly
review the practicality of creating the necessary flow geometry as
demonstrated in water visualization experiments.  In Sec.~4 we review
the actual design and construction features of the experiment.  In
Sec.~5 we develop the all important numerical simulation of the dynamo
growth rate with boundary conditions expected to be achieved in the
experiment.  We conclude in Sec.~6 with the status of the construction
and conclusion.

\section{How the $\alpha\omega$ dynamo works:}\label{sec_dyn}

Fig.~\ref{flux_schem}(a) shows how the $\alpha \omega$ dynamo works in
the New Mexico Dynamo Experiment and Fig.~\ref{flux_schem}(b) shows
how the $\alpha\omega$ dynamo works in the accretion disk forming the
massive, central galactic black hole. In Fig.~\ref{flux_schem}(a)
differential rotation is established in the liquid sodium between two
rotating cylinders as limiting stable Couette flow, $\Omega \propto
1/R^2$, by driving $\Omega_{1} = 4 \Omega_{0}$ where $R_{0} = 2 R_{1}$
and for the disk Fig.~\ref{flux_schem}(b) as Keplerian rotation,
$\Omega \propto 1/ R^{-3/2}$, around the central mass or the black
hole.  This differential rotation wraps up the radial component of an
initial poloidal field Fig.~\ref{flux_schem}(b(A)) either made with
coils or an infinitesimally small, $< 10^{-19} $ G seed field from
density structure at decoupling. The resulting toroidal field becomes
stronger than the initial poloidal field Fig.~\ref{flux_schem}(b(B))
by $B_{toroidal} / B_{poloidal} = n_{\Omega} B_{poloidal}$, where in
equilibrium with resistive decay, the limiting number of turns becomes
$n_{\Omega} \simeq R_{m,\Omega}/2\pi$, and $R_{m,\Omega} = v_0 (R_0 -
R_1)/\eta$ is the magnetic Reynolds number.  This multiplication
factor depends upon the resistivity of liquid sodium or for the disk,
upon the resistivity of the ionized and turbulent plasma. Then a
driven pulsed jet or a collision with the disk by a star,
Fig.~\ref{flux_schem}(b(C))] causes a plume to rise either towards the
end plate or above the disk with the corresponding displaced toroidal
flux forming a loop of toroidal flux.  The radial expansion of the
plume material causes the plane of this loop to untwist or rotate
differentially about its own axis relative to the rotating frame so
that the initial toroidal orientation of the loop is transformed to a
poloidal one, Fig.~\ref{flux_schem}(b(D)).

Resistive diffusion in liquid sodium metal or reconnection in the
ionized plas- \\ ma of the disk allows this now poloidal loop to merge with
the original poloidal field.  For positive dynamo gain, the rate of
addition of poloidal flux must be greater than its decay.  It is only
because the toroidal multiplication can be so large or that
$R_{m,\Omega}$ can be so large that the helicity necessary for gain of
the $\alpha\omega$ dynamo can be much smaller and episodic.

\section{The Water Visualization Experiment:}\label{sec_visual}

The water visualization experiment consists of two parts, the first
establishing the Couette flow and showing that it is stable, and the
second demonstrating the plume-like behavior produced by pulsed jets.
Fig.~\ref{Couette}(A) shows a photograph of the experiment where two
cylinders of plexiglass$^R$ of diameter 15 and 30 cm are rotated at
various speeds of the order of 1 Hz and pulsed jets of water are
injected from a plenum, in turn supplied by pulses of low pressure
air.  Digital video recording of the surface contour is shown from an
angle and from the side in Fig.~\ref{Couette}(B)\&(C), which
demonstrates the hyperbolic contour expected from limiting, maximum
shear, stable Couette flow, $\Omega_1 = \Omega_0 (R_0/R_1)^2$.  The
parabolic profile of the water in solid body rotation within the inner
cylinder is just visible.  Fig.~\ref{plume_sch}(A) shows a schematic of
the apparatus for observing the plumes with a co-rotating camera.
When plumes are injected in stationary flow with an imbedded linear
array or "line" of hydrogen bubbles, Fig.~\ref{plume_sch}(B), one observes
from the side the outline of a rising, diverging vortex, which in turn
simulates a diverging plume.  When the same plume is observed from the
top or axial end and when both the camera and plume are rotating at
$\Omega_0$, then one observes the differential, anticyclonic rotation
of the same imbedded line of bubbles, Fig.~\ref{plume_sch}(C).  In
addition, one observes that the differential rotation of Couette flow
both speeds up the anticyclonic rotation and the dispersal of the
plume.  With these observations, the design of the experiment was
undertaken.

\section{Design of the experiment}\label{sec_design}

The experiment consists of two coaxial cylinders rotating with
different angular velocities, $\Omega_0$ at $R_0$, the outer radius
and $\Omega_1$ at $R_1$, the inner radius, where $R_0 / R_1 = 2$.  The
space between cylinders will be filled with liquid sodium with a small
"topping" of mineral oil. The volume of sodium between the cylinders
is limited by two end plates. One of the plates is solid while the
other plate (referred to as the port plate) has two circular openings
symmetric with respect to the rotation axis of the apparatus and of
diameter $R_{port} = (1/3) R_0$. These ports are in turn connected to
a plenum, supplied with pulsed pressurized liquid sodium, that forms
the pulsed jets.  There is also an annular space between the outer
cylinder and the plenum. These periodic pulses of sodium flow are
driven through the ports by a piston inside the plenum. Sodium flows
out of the circular ports and returns back to the plenum volume
through the annular space between the plenum cylinder and the outer
cylinder. During injection of the jets, the resulting plume
expands. Due to the Coriolis force acting on the expanding plume in
the rotating frame, the plume rotates in the direction opposite to the
direction of the rotation of the vessel in the same fashion as shown
in Fig.~\ref{plume_sch}(C). Such a rotating motion of the plume in the
rotating frame and its axial translation corresponds to an unwinding
helical motion.  This helical motion produces the poloidal magnetic
flux out of toroidal magnetic flux or $\alpha$-effect. The design
drawings of the experiment are shown in Fig.~\ref{exp_phase1} \&
Fig.~\ref{exp_phase2}.

In both figures the magneto active volume with fluid sodium is the
annular space between the differentially rotating inner and outer
cylinders and between the left hand end plate and the middle port
plate. Figure \ref{exp_phase1} is the first stage, the $\Omega$-Phase
of the experiment and presently under construction.  The plume
generating mechanism or $\alpha$-Phase will be added in the second
stage, Fig.~\ref{exp_phase2}.  The purpose of first stage, without the
plumes, is to demonstrate the production of the toroidal field from
the velocity shear and the applied poloidal field.  In addition some
experiments will be performed to investigate the magneto rotation
instability (MRI) using an axial applied field rather that a poloidal
field with a radial component.

The second stage of the experiment will add the drive mechanism to
produce the plumes and possibly lead to positive dynamo gain.  Here
the solid mid-plate has two ports for producing the pulsed jets from
pressure generated in the plenum by a driven piston on the right. The
port plate, plenum and drive mechanism are rigidly connected and
rotate at $\Omega_0$.  The inner cylinder will be rotating faster,
$\Omega_1$, than the outer cylinder in order to create the
differential rotation.  It is driven by its own high speed shaft.  The
secondary drive shaft, gears, belts and motor are not shown. Initially
a 50 kW motor, pulsed to 100 kW will be used.  Greater power can be
applied if needed, but the basic limitation is the mechanical strength
of the outer cylinder which contains the centrifugal pressure of the
sodium.  We have used aluminum for the construction contrary to the
usual practice of stainless steel in reactor coolant technology.  Here
the temperatures are very much less and water vapor is excluded by
mineral oil.  In addition the useful experimental life of the
apparatus is short such that the aluminum corrosion by NaOH will be
negligible.

  Since the end walls rotate at a different rate from the sodium in
Couette flow, the velocity shear at the walls would produce eddy
currents between the sodium and the conducting aluminum walls whenever
an axial field component penetrates both.  Thus in this case, no
conduction is desired between the end walls and sodium and so this
interface will be insulated.  This condition particularly applies when
the field is made purely axial as for the measurements of the magneto
rotational instability, or MRI measurements. On the other hand,
electrical conduction between the sodium and the inner and outer
conducting aluminum cylinders is desired in order to maximize the
magnetic shear for the production of the enhanced toroidal field. Finally
we recognize that the Ekman layer flow at the end walls, a fast radial
flow in a thickness
$\delta_{Eckman}
\simeq R_0 Rey^{-1/2}$, is so thin, $\delta_{Eckman} \simeq 3 \times
10^{-4}R_0 $, that its electrical conduction current is negligable
compared to the primary currents induced by the Couette flow.

The sodium will be heated and liquefied by the hot mineral oil driven
by a recirculating system through the space inside the inner
cylinder. This oil flow also serves to maintain the thermal balance of
the sodium, slightly above the melting temperature and also to prevent
the further heating of the sodium due to the friction heat produced
primarily in the Ekman layers. In addition, the non-recirculating
mineral oil, used to ``top'' the sodium metal in the apparatus, isolates
the liquid sodium from the rotating seals and the one internal
bearing.  This isolation of liquid sodium from the seals and bearing
takes place because the density of oil is $0.86$ of the density of the
sodium at $110^{\circ}\,\mbox{C}$, and so the oil will float to the
central axis of the rotating device. The resulting oil coating will
also ensure the isolation of liquid sodium from the air in the case of
a minor spill as previous experience has shown, \cite{colgate60}.

The maximum shear is desired in the rotational flow between the
cylinders, yet maintaining stable flow.  Therefore, in order to
maximize the shear and maintain minimum fluid drag, or minimize the
torque with the walls and hence, power, one should use Couette flow at
the margin of stability, i.e. when $\Omega_1 R_1^2=\Omega_0
R_0^2$. The New Mexico Dynamo Experiment is designed to have this
marginally stable ratio of angular velocities of the
cylinders. Namely, $R_0/R_1=2$ and $\Omega_1/\Omega_0=4$. In the case
of marginally stable Couette flow the angular velocity profile becomes
\begin{equation}
\Omega=\frac{\Omega_1 R_1^2}{r^2}=\frac{\Omega_0 R_0^2}{r^2}\label{setup_eqn2}
\mbox{.}
\end{equation}

The geometrical parameters of the experiment are: inside radius of the
outer cylinder is $R_0=30.5\,\mbox{cm}$, wall thickness of the outer
cylinder is $\Delta R = 3.2\,\mbox{cm}$, length of the test-volume is
$L=30.5\,\mbox{cm}$, wall thickness of the port and end plates is
$\Delta L=3.2\,\mbox{cm}$, radius of the inner cylinder is
$R_1=15.25\,\mbox{cm}$, the length of the space filled with liquid
sodium in the plenum behind the port plate is $L_1=35.6\,\mbox{cm}$,
radius of the plume ports is $r_p=4.9\,\mbox{cm}$, the radial distance
from the center of the plume port to the rotation axis of the cylinder
is $r_0=R_1+r_p=20.15\,\mbox{cm}$, the width of the annular space
between the outer cylinder and the plenum is $s=2.6\,\mbox{cm}$. One
desires a size as large as possible, but is limited by the costs
implied by available standard bearings, drive belts, material handling
and machine tools.

The material for both cylinders, port and end plates, and the two
ported reservoir plenum cylinders is aluminum alloy 5083-H3. This
alloy has the necessary strength to sustain the centrifugal pressure
of rotating sodium at the required temperatures.  It is widely used in
industry and properties well known. The high- and low-speed drive
shafts, left and right flanges are made of steel. The kinematic
viscosity coefficient of liquid sodium at $110^{\circ}\,\mbox{C}$ is
$\nu=7.1\cdot 10^{-3}\,\mbox{cm}^2\,\mbox{s}^{-1}$. The magnetic
diffusivity of liquid sodium is $\eta=810\,\mbox{cm}^2\,
\mbox{s}^{-1}$, the magnetic diffusivity of the aluminum alloy walls
is $\eta_{Al}=650\,\mbox{cm}^2\,\mbox{s}^{-1}$. 

The analysis of stresses and energy dissipation in the experiment
indicates that the maximum frequency of rotation of the outer cylinder
is limited by the yield strength due to the centrifugal pressure and
is $\leq 33\,\mbox{Hz}$.  This corresponds to $\Omega_0 \leq
207\,\mbox{s}^{-1}$. The angular velocity of the inner cylinder is
always four times larger than the rest of the device and is limited by
$\Omega_1\leq 828\,\mbox{s}^{-1}$. At this angular velocity the
average wall stress in the outer wall, $\Delta R$ at $R_0$ is $\sim
1/4$ of the yield strength.

We define the global hydrodynamic Reynolds number of the Couette flow
in the experiment as
\begin{equation}
\mbox{Re}_{\Omega}=\frac{\Omega_0
R_0(R_0-R_1)}{\nu}\label{setup_eqn4}\mbox{.}
\end{equation} The magnetic Reynolds number for the rotational Couette
flow is defined  in a similar way as
\begin{equation}
\mbox{Rm}_{\Omega}=\frac{\Omega_0 R_0(R_0-R_1)}{\eta}
\label{setup_eqn5}\mbox{.}
\end{equation}
Then, the ratio
$\mbox{Re}_{\Omega}/\mbox{Rm}_{\Omega}=\nu/\eta=\mbox{Pm}$ is a
magnetic Prandtl number. For sodium at $110^{\circ}\,\mbox{C}$ one has
$\mbox{Pm}=8.8\cdot 10^{-6}$. The maximum Reynolds numbers
corresponding to the maximum possible frequency of rotation are
$\mbox{Re}_{\Omega}\approx 1.3\cdot 10^7$ and
$\mbox{Rm}_{\Omega}\approx 120$. This magnetic Reynolds number is
higher than the one achieved in the successful dynamo experiments
carried out so far, and we hope that this will allow us to investigate
a wider spectrum of the behavior of conducting liquids. Particularly,
the observations of MHD turbulence may be possible. At the same time,
slower rotation rates can produce flows with as low magnetic Reynolds
number as desired.

The experiment will also have current coils to produce an external
magnetic field within the magneto active volume.  These fields will be
both with primarily radial (poloidal) or primarily axial components
depending upon the emphasis of the the experiment, i.e., toroidal gain
or MRI.  All three components of the magnetic field will be measured
by miniature Hall-effect detectors placed at a various radii inside an
aerodynamically shaped probe inside the sodium.  In the second stage,
with driven plumes, we hope to measure the flux and its orientation
transported by the plumes or $\alpha$ effect.  These measurements will
be performed by an array of magnetic detectors placed on the surface
of the end plate opposite a plume. The fast response of the
Hall-effect detectors, micro seconds, will allow one to record the
detailed time evolution of the magnetic field produced by a single
plume.  A radial array of 5 pressure transducers will allow an
accurate measurement of the pressure profile and thus a measurement of
the Couette profile.


\section{Numerical simulations of the dynamo}\label{sec_numerics}
\newcommand{\ptl}{\partial}

At the current stage of numerical modeling of the New Mexico Dynamo
Experiment we assumed the following simplified model. We consider the
flow of the liquid sodium only inside the cylindrical, annular, space
bounded by the radii $R_1$ and $R_0$ and the end plate and port plate
at $z=L/2$ and $z=-L/2$.  The walls are assumed perfectly
conducting. In view of the conductivity of aluminum walls being higher
than the sodium, this assumption qualitatively predicts the evolution
of magnetic fields and the growth rate of the dynamo. However, to
obtain more accurate results, one needs to consider more realistic
boundary conditions taking into account the finite thickness of the
walls, the insulating air, i.e., vacuum, outside the device, and the
plenum filled with liquid sodium at one end. We also assume a
simplified kinematic model for the flow and do not actually solve the
hydrodynamic equations. In other words we do not take into account the
pondermotive force (${\bf j}\times{\bf B}$).

The kinematic dynamo equation is
\begin{equation}
\frac{\ptl {\bf B}}{\ptl t}=-\nabla\times\left(\eta\nabla\times{\bf B}\right)
+\nabla\times({\bf v}\times{\bf B})\label{eqn_num5}\mbox{,}
\end{equation}
where $\eta$ is the a coefficient of magnetic diffusivity,
$\displaystyle\eta=\frac{c^2}{4\pi\sigma}$. Instead of solving the
dynamo equation for ${\bf B}$ we introduce potentials ${\bf A}$
and $\varphi$. Using the gauge condition
$c\varphi-{\bf v}\cdot{\bf A}+\eta\nabla\cdot{\bf A}=0$ one can
derive the following equation for the evolution of the vector potential
${\bf A}$
\begin{equation}
\frac{\ptl A^i}{\ptl t}=-A_k \frac{\ptl v^k}{\ptl x^i}- v^k\frac{\ptl A^i}
{\ptl x^k}+\eta\frac{\ptl^2 A^i}{\ptl x^k \ptl x_k}\label{eqn_num6}\mbox{,}
\end{equation}
where the magnetic diffusivity $\eta$ is assumed to be constant
throughout the cylinder and the coordinate notations refer to a
Cartesian coordinate system $x^i$. We use our 3D kinematic dynamo code
to solve Eq.~(\ref{eqn_num6}).  Then, the magnetic field can be
obtained at any time by taking the curl of ${\bf A}$. The code is
written in cylindrical coordinates and uses an explicit scheme with
the central spatial differencing in the advection term and the
standard nine points stencil for the diffusion term. We assume all
boundaries of the cylinder ($r=R$, $z=-L/2$, and $z=L/2$) to be
perfect conductors. Then, the boundary conditions for ${\bf A}$ at
perfectly conducting boundaries compatible with the gauge can be
chosen as follows: the components of ${\bf A}$ parallel to the
boundary are zeros and the divergence of ${\bf A}$ at the boundaries
is zero.  This gives three boundary conditions for three components of
the vector potential.

We use the following dimensionless units: the unit of length is the
outer radius of the test volume, $R_0$, the unit of velocity is the
azimuthal velocity of the outer cylinder, $\Omega_0 R_0$. Therefore,
the outer cylinder makes one revolution during the dimensionless time
$2\pi$, the inner cylinder makes one revolution during the
dimensionless time $\pi/2$.  The critical Couette velocity profile is
assumed, $\Omega=1/r^2$ in our units. The inner radius is $1/2$ in our
units, the length of the test volume is $1$. In order to account for
the conducting material inside the inner cylinder we extended our
computational region toward $r=0.5$ and assumed solid body rotation
for $r<0.2$ with the angular velocity equal to the angular velocity of
the inner cylindrical wall of the test volume. We work in cylindrical
coordinates with the axis of rotation centered on the axis of symmetry
of the device. The $z=-0.5$ plane is the port plate of the test
volume, the $z=0.5$ plane is the end plate of the test volume.

We assume that there is a bias magnetic field produced by external
coils and frozen in the ideally conducting boundaries. To approximate
this field we choose the initial conditions in the form $A_r=0$,
$A_z=0$, $A_{\phi}=r(z+0.5)$.  Then, the initial magnetic field is
$B_r=-r$, $B_z=2(z+0.5)$, $B_{\phi}=0$. This poloidal field satisfies
the equation $\nabla^2 {\bf B}=0$ and, therefore, is unchanged if only
rotation is present.

First, we do simulations without plumes, when the flow velocity inside
the device is Couette flow given by Eq.~(\ref{setup_eqn2}).  Both the
poloidal and stationary state toroidal fields are shown in
Fig.~\ref{f_numexp1} for $\mbox{Rm}_{\Omega}=120$.  The ratio of the
toroidal field to the poloidal field is $\sim 20$ and depends on the
position of measurement inside the test volume. The toroidal field
near the end plate has the opposite sign from the toroidal field in
the middle of the test volume. The reversal of the sign of the
toroidal magnetic field can be understood in terms of the conservation
of total flux of the toroidal magnetic field through the cross section
of the cylindrical computational space. Since this space is bounded by
an ideal conductor, the total magnetic flux cannot change. Initially,
the toroidal magnetic field was zero, so the total net flux of the
toroidal magnetic field should remain zero. Therefore, regions of the
magnetic field with different signs must exist in the rotating liquid.
Of course, there is no actual discontinuity of the fluid velocity near
the end plates of the test volume. Instead, the Eckman boundary layers
develop at the end plates. However, the approximation of the Eckman
layer by a mathematical discontinuity of the toroidal velocity near
the end plates has very little effect on the structure of the magnetic
fields excited in the conducting rotating fluid under the conditions
of the experiment.

Next, we modeled the kinematic dynamo produced by jets of sodium
together with the Couette differential rotation.  The plume flow is
interposed onto a background Couette differential rotation occupying
the whole computational domain $\displaystyle {\bf v}_c
=\frac{1}{r}{\bf e}_{\phi}$.  A jet of liquid sodium is simulated by a
vertically progressing cylinder of radius $r_p$ in the co-rotation
frame that starts at the port plate of the disk located at $z=-L/2$
and emerges to a height of $z=L/2$. At the same time the cylinder axis
moves with the local Couette velocity. By the time the plume reaches
its highest point the jet cylinder rotates by $\pi$ radians.  The
cylinder does not rotate with respect to the rest frame. Therefore, in
the frame rotating with the Keplerian velocity, it untwists by $\pi$
radians during the time it rises. At the same time the axis of the
cylinder moves by $\pi$ radians around the central axis participating
in the rotation with the Couette speed.  The length of the cylinder
increases with time and its velocity, $v_{pz} \approx v_c$.  The
vertical velocity of the liquid inside the cylinder is constant and is
equal to $v_{pz}$.  After the time the plume rotates by $\pi$ it is
stopped and the velocity field is restored to be pure Couette
differential rotation everywhere.  This very simplified flow field
captures the basic features of the actual and complicated flow
produced by the driven sodium jets. We note, however, that in view of the
water visualization experiments, Fig.~\ref{Couette} and
Fig.~\ref{plume_sch}, that this approximation of constant radius plumes
may be a conservative approximation, because the actural plumes  diverge
in radius leading to a progressive increase in $R_{m,plume}$.

Since Eq.~(\ref{eqn_num6}) requires spatial derivatives of the
velocity components, we apply smoothing of discontinuities in the flow
field described above. Also we introduce smooth switching on and off
of the jets in time. For all three components of velocity $v^k$ we use
the same interpolation rule, which for two plumes is
\begin{equation}
v^k=v^k_{in1}s_1+v^k_{in2}s_2+(1-s_1-s_2)v^k_{out}\mbox{.}\label{eqn_num7}
\end{equation}
Here $s_1(r,\phi,z,t)$ and $s_2(r,\phi,z,t)$
are smoothing functions for plume $1$ and $2$ correspondingly. Each function
$s$ is close to $1$ in the region
of space and time occupied by the plume and is close to $0$ in the
rest of space and during times when the plume is off. Transition from
$1$ to $0$ happen in a narrow layer at the boundary of the plume and
during the interval of time short compared to the characteristic time
of the plume rise. $v^k_{in1}$ and $v^k_{in2}$ are velocities of the
flow of plumes $1$ and $2$, $v^k_{out}$ is the velocity of the flow
outside the regions occupied by the plumes.
For spatial derivatives of the velocity components, one
has from expression~(\ref{eqn_num7})
\begin{eqnarray}
&& \frac{\ptl v^k}{\ptl x^i}=\frac{\ptl s_1}{\ptl x^i}\left(v^k_{in1}-
v^k_{out}\right)+\frac{\ptl s_2}{\ptl x^i}\left(v^k_{in2}-v^k_{out}
\right)+ \nonumber\\
&& s_1\frac{\ptl v^k_{in1}}{\ptl x^i}+s_2\frac{\ptl v^k_{in2}}{\ptl x^i}
+(1-s_1-s_2)\frac{\ptl v^k_{out}}{\ptl x^i}\mbox{.}\label{eqn_num8}
\end{eqnarray}
It is easy to extend this approach for the case of arbitrary number of plumes.

Let us assume that the cylindrical jet, going in the positive
direction of the $z$ axis, is launched at the position of the axis of
the jet at $r=r_0$ and $\phi=\phi_0$.  Let us denote this plume as
number $1$ and the symmetric plume also ejected at the same time as
number $2$.  Then, after time $(t-t_p)$ from the starting moment of
the plume $t=t_p$, its position is
\begin{equation}
\phi_1=\phi_0+(t-t_p)r_0\Omega_{c0}\mbox{,}\label{eqn_num9}
\end{equation}
where $\Omega_{c0}=\Omega_c(r_0)$ is the Couette angular
rotational velocity at $r=r_0$.
The position of the axis of the symmetric plume is
\begin{equation}
\phi_2=\phi_1+\pi\mbox{.}\label{eqn_num10}
\end{equation}
The radii of both plumes are $r_p$. The originating surface of the
plumes $1$ and $2$ is at the port plate at $z=-L/2$. The leading
surface of the plume $1$ is at $z_1=-L/2+v_{pz} (t-t_p)$ and the
leading surface of the plume $2$ is $z_2=z_1$. The velocity field
inside the upward jet number $1$ is
\begin{eqnarray}
&& v^r_1=r_0\Omega_{c0}\sin (\phi-\phi_1)\mbox{,}\label{eqn_num11} \\
&& v^{\phi}_1=r_0\Omega_{c0}\cos (\phi-\phi_1)\mbox{,}\label{eqn_num12} \\
&& v^z_1=v_{pz}\mbox{.}\label{eqn_num13}
\end{eqnarray}
The same velocity field given by expressions~(\ref{eqn_num11}--\ref{eqn_num13})
is inside the jet number $2$ with the obvious replacement
of $\phi_1$ by $\phi_2$.
We choose the following interpolation functions
\begin{equation}
s_1=\left(\frac{1}{2}+\frac{1}{\pi}\arctan\frac{r_p^2-{r^{\prime}_1}^2}
{2r_p\Delta}\right)
\left(\frac{1}{2}+\frac{1}{\pi}\arctan\frac{(z+L/2)(z_1-z)}{
\Delta\sqrt{(z_1+L/2)^2+\Delta^2}}\right)S(t)\label{eqn_num14}
\end{equation}
and
\begin{equation}
s_2=\left(\frac{1}{2}+\frac{1}{\pi}\arctan\frac{r_p^2-{r^{\prime}_2}^2}
{2r_p\Delta}\right)
\left(\frac{1}{2}+\frac{1}{\pi}\arctan\frac{(z+L/2)(z_2-z)}{
\Delta\sqrt{(z_2+L/2)^2+\Delta^2}}\right)S(t)\mbox{.}\label{eqn_num15}
\end{equation}
Here ${r^{\prime}_1}^2=r_0^2+r^2-2r_0r\cos(\phi-\phi_1)$ is the distance
from the axis of the plume $1$,
${r^{\prime}_2}^2=r_0^2+r^2-2r_0r\cos(\phi-\phi_2)$ is the distance
from the axis of the plume $2$, $\Delta$ is the thickness of the transition
layer of the functions $s_1$ and $s_2$ from their value $1$ inside the
plume to $0$ outside the plume, $\Delta \ll r_p$.
The square root expressions in $z$--parts of
$s_1$ and $s_2$ ensure that the thickness of the transition layer in
the
$z$ direction is never less than $\Delta$, even just after the plumes
are started, when the differences $(z_1+L/2)$ and $(z_2+L/2)$ are
zero. We choose $\Delta=0.01$.

The function $S(t)$ ensures a smooth ``turning on'' and ``turning
off'' of the plumes at prescribed moments of time. If the plumes are
to be started at $t=t_p$ and to be terminated at $t=t_d$ ($t_d>t_p$),
then we adopt the following form of the function $S(t)$
$$
\left\{
\begin{array}{ll}
S(t)=0\mbox{,} \quad & \mbox{for} \quad t<t_p-\delta t/2\mbox{,}  \\
S(t)=\frac{1}{2}+\frac{1}{2}\sin\left(\pi\frac{t-t_p}{\delta t}\right)
\mbox{,}
\quad & \mbox{for} \quad t_p-\delta t/2 < t < t_p + \delta t/2 \mbox{,} \\
S(t)=1\mbox{,} \quad &
\mbox{for} \quad t_p+ \delta t/2 < t < t_d-\delta t/2\mbox{,}\\
S(t)=\frac{1}{2}-\frac{1}{2}\sin\left(\pi\frac{t-t_d}{\delta t}\right)\mbox{,}
\quad & \mbox{for} \quad t_d-\delta t/2 < t < t_d + \delta t/2 \mbox{,} \\
S(t)=0\mbox{,} \quad & \mbox{for} \quad t>t_d+\delta t/2\mbox{.}
\end{array}
\right.
$$
where $\delta t$ is the length of the transition period. $S=0$ corresponds
to the flow without plumes, $S=1$ corresponds to the flow with plumes.
One needs to ensure that $\delta t < t_d-t_p$. We took
$\delta t = (t_d-t_p)/5$. The cycles with the cylindrical jets present are
interchanged periodically
with the cycles without such jets and with the pure Couette
rotation only. The time between two subsequent ejections is $\Delta t_p$ and
we always have $\Delta t_p > t_d-t_p$, such that at any time only one
pair of plumes are present. 

We take the radius of the jet $r_p=0.21$, the position of the center
of the jet at $r_0=0.71$, and the vertical velocity of the flow inside
the rising cylinder $v_{pz}=0.63$. This geometry corresponds to the
experimental setup described in Sec.~\ref{sec_design}.  The radius
of the jet is chosen to be the maximum possible, and the vertical
velocity of the liquid in the jet requires only moderate power of the
piston driving mechanism. The vertical velocity of the plume is chosen
such that the plume reaches the end plate during the time when the
fluid at the radius of the plume rotates by the angle $\pi$.  During
this same time the plume rotates clockwise by $\approx \pi$ radians in
the local Couette rotating frame. Such a timing should maximize the
release of the poloidal magnetic flux due to the diffusion out of the
twisting plume into surrounding conducting fluid.  After the plume
reaches the end plate, the velocity field of the flow is smoothly set
back to the pure Couette profile without further poloidal
motions. There are two identical plumes ejected simultaneously through
the two orifices located symmetrically with respect to the rotation
axis of the device. We performed simulations with different time
intervals between subsequent plume ejections. We looked at the various
rates of: (1) one pair of plumes per one revolution of the outer
cylinder (i.e., per $2\pi$ units of time), (2) one pair of plumes per
two revolutions of the outer cylinder, (3) one pair of plumes per
three revolutions of the outer cylinder, and so on, until the dynamo
could no longer be excited. The typical curve of the energy growth of
the magnetic field is presented in Fig.~\ref{f_numexp2}. The
dependence of the growth rate of the dynamo vs. the rate of plumes
(the rate is the inverse of the number of revolutions of the outer
cylinder, $N$, per one ejected pair of plumes) is shown in
Fig.~\ref{f_numexp4}.  One can see that the threshold for the dynamo
excitation is somewhere between one pair of plumes per 4 revolutions
and one pair of plumes per 5 revolutions. The smallest point on the
graph for $N=5$ may not have converged numerically, and so is likely
to be below the axis where $\gamma=0$.


\section{Present Status}\label{sec_present}

The first phase of the experiment, the $\Omega$-Phase, is largely
completed, Fig.~\ref{construct}.  The electronics are still a major
consideration where the data from roughly 120 detectors, magnetic,
pressure and temperature, must be transmitted digitally to a computer
from the rotating equipment.


\section{Conclusions}\label{sec_concl}

The numerical simulations along with the engineering design
feasibility give us confidence that, with a continuing effort a
positive gain $\alpha\omega$ dynamo can be made in the laboratory.

\section*{Acknowledgements}

This work was partially supported by the US Department of Energy,
under contract W-7405-ENG-36 and NSF Grant 9900661.  We particularly
acknowledge the initial  support of the New Mexico Institute of Mining
and Technology, (NMIMT)  through the Energetic Materials Research and
Testing Center (EMRTC) that allowed this project to begin.

We would like to express our gratitude to J.~M.~Finn, LANL for numerous
discussions on magnetic dynamos and help with the numerical code and
Hui Li, LANL, for discussions on how this work joins to astrophysics.

\newcommand{\noopsort}[1]{}

\received{\today}
\clearpage


\begin{figure}
\caption[The Magnetic Flux Distribution of the $\alpha\omega$ 
Dynamo]{A schematic of
the sodium dynamo experiment Fig.~1(a) in comparison to the accretion 
disk dynamo of
Fig.~1(b). In both cases differential rotation in a conducting fluid 
wraps up an
initial radial field component into a much stronger toroidal field. Then
either liquid sodium plumes driven by  pulsed jets or star-disk
collisions eject and rotate  loops of toroidal flux into the
poloidal plane. Resistivity or reconnection merges this new or
additional poloidal flux with the original poloidal flux leading
to dynamo gain.}
\label{flux_schem}
\end{figure}

\begin{figure}
\caption[The Water Visualization Experiment]{(A) The plume rotation experiment
apparatus with its camera mounts, electrical communications,
air supply, and the cylinder drive systems. (B) The surface  of the
maximum shear stable Couette flow.  Note the smooth hyperbolic surface
indicative of low turbulence.  (C) The side view of the contour of the
maximum shear, stable Couette flow. Note the hyperbolic surface external to
the inner  cylinder and the parabolic surface of the water inside the
inner cylinder in solid body rotation.}
\label{Couette}
\end{figure}

\begin{figure}
\caption[The Plume Rotation Experiment] {(A) Schematic of the  plume rotation
experiment apparatus for viewing the rising plumes from the side and the plume
rotation from above. (B) The entrained small bubbles of hydrogen from
pulsed electrolysis outlines the rising plume without rotation from the
pulsed jet. (C) The same plumes viewed from above in a rotating frame.
The line of bubbles trace a progressive rotation of the plume as a
function of rotation of the frame. }
\label{plume_sch}
\end{figure}

\begin{figure}
\caption[The first or $\Omega$-phase of the Experiment]{The detailed 
design drawing
of the rotating components of the $\Omega$-Phase of the New Mexico 
Dynamo Experiment.
By comparing  Figs.~\protect\ref{exp_phase1} 
and~\protect\ref{exp_phase2} the labeling
of the various components can be compared and identified in the two
drawings. The main cylinder of radius $R_1=30.5\,\mbox{cm}$ rotates between
two bearing mounts. In the $\Omega$-Phase there is no  hydraulic 
drive to produce the
plumes even though the constructed  aluminum parts of
Fig.~\protect\ref{exp_phase2} show the port plate and  two ported 
reservoir plenum
cylinders. These parts are necessary to  support the port plate which 
defines one
end of the Couette flow annular experimental volume.}
\label{exp_phase1}
\end{figure}

\begin{figure}
\caption[The second or $\alpha$-phase of the Experiment]{The design 
drawing of both the
rotating components as well  as the plume drive mechanism of both the 
$\alpha$- and
$\Omega$-Phases of the New Mexico Dynamo Experiment.}
\label{exp_phase2}
\end{figure}

\begin{figure}
\caption[The Poloidal and Toroidal Magnetic Fields in the $\Omega$ Phase
of the Experiment]{The top panel shows the poloidal magnetic field
and the bottom panel shows
the contours of equal values of the toroidal magnetic field.
The toroidal magnetic field is shown after the steady state
between stretching and diffusion of the magnetic field has been reached.
\label{f_numexp1} }
\end{figure}

\begin{figure}
\caption[The Time Evolution of the Magnetic Energy of the Dynamo]{Part of one
simulation for one pair of plumes per three revolutions showing 
exponential growth of
the dynamo. The small spikes on the graph are due to the action of 
each single plume.
The slowly rising and decaying arches are due to the production of the
toroidal field from the fraction of the poloidal field added by the plume.
\label{f_numexp2} }
\end{figure}

\begin{figure}
\caption[The Dependence of the Growth Rate of the Dynamo on the Plume Rate]{
The dependence of the growth rate of the dynamo on the plume rate. $N$ is
the number of revolutions of the outer cylinder from one ejection of a pair
of plumes to the next ejection of a pair of plumes.
\label{f_numexp4} }
\end{figure}

\begin{figure}
\caption[The initial   rotating components of the Liquid
Sodium $\alpha\omega$ Dynamo] {The plume port plate and reservoir
plenum cylinders are shown yet to be mounted internally. The plume
drive piston is not constructed in this phase. The plume port
plate is supported by the  reservoir plenum cylinder for greater
rigidity. The rotating drive components of the experiment are  mounted
on the bearing supports. The massive bearing supports
and base plate   are designed to reduce the rotating
vibration amplitude.  The oscillating hydraulic drive mechanism for the
jet--plume production is not constructed, awaiting the $\alpha$-Phase
of the experiment.}
\label{construct}
\end{figure}

\end{document}